\documentstyle[11pt,palatino]{article}
\def\beq{\begin{equation}}
\def\eeq{\end{equation}}
\def\bea{\begin{eqnarray}}
\def\eea{\end{eqnarray}}
\def\bq{\begin{quote}}
\def\eq{\end{quote}}

\def\bq{\begin{quote}}
\def\eq{\end{quote}}

\def\mpl{\ifmmode \overline M_{P}\else $\overline M_{P}$\fi}
\begin{document}
\renewcommand{\thefootnote}{\fnsymbol{footnote}}
\begin{flushright}
CERN-TH/2001-323\\
SINP/TNP/01-27\\
hep-ph/0111345 \\
\end{flushright}
\vskip 5mm
\begin{center}
{\Large \bf Testing orbifold models of Supersymmetric Grand Unification}\\
\vspace*{1cm}
{\large\bf Gautam Bhattacharyya ${}^1$\footnote{gb@theory.saha.ernet.in}
          {\rm and} K.~Sridhar ${}^2$\footnote{sridhar@theory.tifr.res.in}
}\footnote{On leave of absence from the Tata Institute of Fundamental
Research, Homi Bhabha Road, Mumbai 400 005, India.}\\
\vspace{10pt}
{\sf 1. Saha Institute of Nuclear Physics, 1/AF Bidhan
        Nagar, Kolkata 700064, India \\
     2. Theory Division, CERN, CH-1211, Geneva 23, Switzerland }

\normalsize

\vspace*{2cm}
{\bf Abstract}
\end{center}

In a model of supersymmetric SU(5) Grand Unification with a spatial
dimension described by the orbifold $S^1/(Z_2 \times Z_2')$, proton
decay is naturally suppressed at all orders.  This is achieved by a
suitable implementation of the discrete symmetries on the brane. But
baryon number violating interactions are present in this model. We
propose a few possible experimental tests of this model which exploit
the effect of the baryon number violating couplings on low-energy
observables like neutron-antineutron oscillations, double nucleon
decay into two kaons, hadronic decay widths of the $Z$ boson, and $t
\bar t$ production cross-section in Run II of the Tevatron.

\noindent

\vspace*{1cm}
\noindent
\setcounter{footnote}{0}
\renewcommand{\thefootnote}{\arabic{footnote}}

\vfill
\clearpage

\setcounter{page}{1}
\pagestyle{plain}

\noindent
Weak scale Supersymmetry (SUSY) is arguably the most promising
candidate for describing physics beyond the Standard Model (SM). It
was introduced in the first place to protect the electroweak scale
from destabilizing divergences \cite{susy}. Later, the idea of SUSY
being embedded into some Grand Unified Theories (GUT) \cite{susygut}
received a significant boost when it was observed that the Minimal
Supersymmetric Standard Model (MSSM) has the right particle content to
ensure that the gauge couplings indeed meet a high scale $M_{\rm GUT}
\sim 2 \times 10^{16}$ GeV \cite{unifi}. The unification condition, in
fact, predicts the existence of superparticles in the TeV range which
suggests that they may well be discovered in collider experiments in
the near future.

While the SUSY GUT models are theoretically very appealing, they face
stiff experimental challenges. For instance, the most minimal version
of the supersymmetric SU(5) model has now been excluded
\cite{murayama} by the Super-Kamiokande lower limit on the proton
lifetime in the $p \to K^+ \bar{\nu}$ channel ($6.7 \times 10^{32}$
years at 90\% C.L. \cite{sk}). The argument goes as follows: Requiring
that the gauge coupling constants exactly unify at the GUT scale, a
precise knowledge of the strong coupling constant $\alpha_S(M_Z)$ from
LEP has now sharpened the interval $3.5 \times 10^{14} \leq M_{H_C}
\leq 3.6 \times 10^{15}$ (in GeV) at 90\% C.L., where $M_{H_C}$ is the
mass of the coloured triplet Higgs boson that mediates the dominant
decay of the proton. On the other hand, the new Super-Kamiokande limit
mentioned above imposes the constraint: $M_{H_C} \geq 7.6 \times
10^{16}$ GeV. Such a gross disagreement between these two limits
excludes minimal supersymmetric SU(5) even in the decoupling limit. To
save the GUT models from this embarassment, one can either look for
suppressing the dimension-5 proton decay operator in some way, or
attempt to prepare a natural ground to solve the doublet-triplet
splitting problem by pushing the coloured triplet Higgs mass very
heavy while keeping the doublet Higgs at the weak scale.

In recent years, there has been a great resurgence of interest in the
physics of extra dimensions. More recently, the implications of models
in which SUSY is embedded into GUT in higher dimensions, and both SUSY
and GUT breaking being eventually realised by orbifold
compactifications, have been explored. There are several models in the
literature \cite{kawamura,altfer,koba}, but one recent suggestion,
originally due to Kawamura \cite{kawamura} and developed, in
particular, by Altarelli and Feruglio \cite{altfer}, is particularly
interesting because it provides a framework for an explicit
realisation of minimal SUSY embedded into SU(5) GUT while avoiding all
the aforementioned pitfalls of conventional SUSY GUT scenarios. We
refer to this model as the Kawamura-Altarelli-Feruglio (KAF) model. We
would like to point out at the very outset, that the model of
Altarelli and Feruglio \cite{altfer} does differ in the details of the
interaction of the brane and bulk fields from the original Kawamura
framework \cite{kawamura} and in all such cases of difference what we
refer to as the KAF model is only the Altarelli-Feruglio model.

To recapitulate the basics, the minimal fermionic representation of
the 5-dimensional Lorentz group is a 4-component spinor, which
decomposes under the 4-dimensional Lorentz group as two 2-component
Weyl spinors. Thus in 5 dimensions the minimal SUSY has 8 real
supercharges which corresponds to $N = (1,1)$ or equivalently $N = 2$
supersymmetry. In other words, a hypermultiplet ($\Phi(x^\mu,y),
\Phi^c(x^\mu,y)$) in 5 dimensions corresponds to two sets of
degenerate $N = 1$ chiral multiplets $\Phi(x^\mu,y)$ and
$\Phi^c(x^\mu,y)$ in 4-dimensional language with the 5th ($y$)
coordinate acting as a label on the 4-dimensional fields. The
5-dimensional theory is necessarily a vector-like theory with $\Phi$
and $\Phi^c$ transforming under SU(5) as 5 and $\overline{5}$,
respectively. Also, in 5 dimensions the only interaction is the gauge
interaction.  In the KAF model, one starts with a 5-dimensional GUT
with minimal SU(5) gauge group and $N = 2$ SUSY. The 5-dimensional
spacetime is factorised into a product of the four dimensional
spacetime $M^4$ (labelled by the co-ordinates $x^{\mu}$) with the
extra spatial dimension compactified on the orbifold $S^1/(Z_2 \times
Z_2')$ (labelled by the co-ordinate $y=x_5$)). The radius, $R$, of the
circle $S^1$ is chosen to be of the order of $M_{\rm GUT}^{-1}$.  The
orbifold construction is as follows: One starts by dividing $S^1$ by a
$Z_2$ transformation $y \rightarrow -y$ and then a further division by
$Z_2'$ which acts as $y' \rightarrow -y'$ with $y'=y+\pi R/2$. After
these identifications, the spacetime is the interval $[0, \pi R/2]$
with a brane located at each fixed point $y=0$ and $y=\pi R/2$. As a
result of the two reflections, the branes at $y=\pi R$ and $-\pi R/2$
are identified with those at $y=0$ and $y=\pi R/2$, respectively. The
reason for the action of two discrete symmetries $Z_2$ and $Z_2'$
becomes apparent when the following considerations are taken into
account. Let us consider a generic field $\phi(x^{\mu}, y)$ exisiting
in the 5-dimensional bulk. The $Z_2$ and $Z_2'$ parities (called $P$
and $P'$, respectively) are defined for this field as

\begin{eqnarray}
\phi(x^{\mu},y) &\rightarrow & \phi(x^{\mu},-y) = P\phi(x^{\mu},y) ,
\nonumber \\
\phi(x^{\mu},y') &\rightarrow & \phi(x^{\mu},-y') = P'\phi(x^{\mu},y') .
\end{eqnarray}

Using the notation $\phi_{\pm\pm}$ for the fields with $(P,P')=(\pm,
\pm)$, a Fourier expansion in $y$ yields:
\begin{eqnarray}
\phi_{++}(x^{\mu},y) &=& \sqrt{{2 \over \pi R}} \sum_{n=0}^{\infty}
\phi_{++}^{(2n)}(x^{\mu}) {\rm cos}{2ny \over R} ,
\nonumber \\
\phi_{+-}(x^{\mu},y) &=& \sqrt{{2 \over \pi R}} \sum_{n=0}^{\infty}
\phi_{+-}^{(2n+1)}(x^{\mu}) {\rm cos}{(2n+1)y \over R} ,
   \\
\phi_{-+}(x^{\mu},y) &=& \sqrt{{2 \over \pi R}} \sum_{n=0}^{\infty}
\phi_{-+}^{(2n+1)}(x^{\mu}) {\rm sin}{(2n+1)y \over R} ,
\nonumber \\
\phi_{--}(x^{\mu},y) &=& \sqrt{{2 \over \pi R}} \sum_{n=0}^{\infty}
\phi_{--}^{(2n+2)}(x^{\mu}) {\rm sin}{(2n+2)y \over R} \nonumber .
\end{eqnarray}

The above Fourier decompositions lead to the following observations:
Upon compactification, the fields $\phi_{++}^{(2n)}$ acquire a mass
$2n/R$, while $\phi_{+-}^{(2n+1)}$ and $\phi_{-+}^{(2n+1)}$ acquire a
mass $(2n+1)/R$ and $\phi_{--}^{(2n+2)}$ acquire a mass
$(2n+2)/R$. This implies that the only fields which can have massless
components are $\phi_{++}^{(2n)}$. The other interesting consequence
is that only $\phi_{++}$ and $\phi_{+-}$ can have non-vanishing
components on the $y=0$ brane. These simple observations have
remarkable consequences. For example, consider the case when
$\phi(x^{\mu},y)$ transforms as a multiplet under some symmetry group
G. Now if $P$ (or $P'$) are chosen to be different for different
components of the multiplet, then upon compactification a symmetry
reduction will result.  The usual line of action is the following:
Start with a 5-dimensional $N = 2$ SUSY theory invariant under the
gauge group SU(5); upon the first compactification by $Z_2$ the
conjugated fields are projected out and the $N = 2$ SUSY reduces to $N
= 1$ SUSY but still respecting the gauge SU(5), on the second
compactification by $Z_2'$ the SU(5) gauge symmetry is broken to the
SM gauge group SU(3) $\times$ SU(2) $\times$ U(1) with a unbroken $N =
1$ SUSY.  The assignment of parities in the KAF model (which we
briefly summarise here but for details refer the readers to the
original references \cite{kawamura, altfer}) are such that only the
(+,+) fields, i.e.  fields which have massless components and do not
vanish on the $y=0$ brane (which is taken to be the `visible' brane)
are the gauge and Higgs multiplets of the MSSM. Other components like
the coloured triplet Higgs fields, for example, do not have (+,+)
parity assignments and therefore do not have any massless component
but acquire a minimum mass of order $1/R \sim M_{\rm GUT}$. This
provides a very elegant solution to the problem of doublet-triplet
splitting and constitutes one of the attractive features of the
orbifold compactification advocated in the KAF model.

The 5-dimensional theory contains the vector bosons $A_M$ ($M =
(\mu,5)$, where $\mu = 0,1,2,3$), two gauginos $\lambda, \lambda^c$
and a real scalar $\sigma$, all of them transforming in the adjoint
representation of SU(5). This is equivalent to a $N = 1$ vector
multiplet $V (A_\mu, \lambda)$ and a chiral multiplet $\Sigma
(\phi_\Sigma, \psi_\Sigma)$ in the adjoint representation, with
$\phi_\Sigma = (\sigma + i A_5)/\sqrt{2}$ and $\psi_\Sigma =
\lambda^c$.  Then there are $N = 1$ chiral Higgs multiplets $H_5$ and
$H_{\overline{5}}$ and their conjugated partners. $H_5$ and
$H_{\overline{5}}$ contain the scalar Higgs doublets $H_u^D$ and
$H_d^D$ and the corresponding scalar triplets $H_u^T$ and
$H_d^T$. Their $P$ and $P'$ parity assignments are listed in Table 1
of \cite{altfer}. We just mention that $A_\mu$, $\lambda$, $H_u^D$ and
$H_d^D$ are (+,+) fields, which have zero modes and also can have
nonvanishing components at the branes.

For the quark and lepton matter, there is more freedom in the
assignments of parities. First of all, we notice that $N = 2$ SUSY
does not allow any trilinear Yukawa interaction in the bulk. In $N =
2$ SUSY, all interactions are gauge interactions. The Yukawa
interactions can however be placed in the branes where $N = 2$ SUSY
breaks to $N = 1$ SUSY. Now, as Altarelli and Feruglio \cite{altfer}
have argued, the matter fields are located only at the branes. They
cannot propagate in the bulk, because if they do, it is not possible
to construct a bulk interaction invariant under the gauge SU(5) and
the $(Z_2, Z_2')$ parities. For the sake of illustration, let us
denote the matter fields in the SU(5) representation as: $10 \equiv
(Q, U, E)$ and $\overline{5} \equiv (L, D)$, which are $N = 1$ chiral
multiplets. Since all these matter fields can reside in the $y = 0$
brane, the $Z_2$ parities for all of them are positive. What about the
$Z_2'$ parities of the matter fields? They have to be fixed from the
gauge interaction of the matter fields with the gauge fields and the
Yukawa interaction of the matter fields with the Higgs fields in 4
dimensions.  In the following, let us briefly summarise how it is
realised in the KAF model.

First we explicitly write down all the relevant interactions. For that
we denote a matter chiral multiplet by $\Phi_M \equiv (\phi_M,
\psi_M)$, where $M = (Q, L, U, D, E)$, with $\phi_M$ and $\psi_M$
being the scalar and fermionic components respectively. The gauge
interaction induced at $y = 0$ by the Kahler potential,
\begin{eqnarray}
K \equiv 10^\dagger e^V 10 + {\overline{5}}^\dagger e^V \overline{5},
\end{eqnarray}
when decomposed in components, reads:
\begin{eqnarray}
{\cal L}_{g}={\cal L}^a_{g}+{\cal L}^{\hat a}_{g}~,
\end{eqnarray}
\begin{eqnarray}
{\cal L}^a_{g}=\sum_M \overline{\psi_M} \bar\sigma^\mu T^a \psi_M A^a_\mu~,
\label{lg1}
\end{eqnarray}
\begin{eqnarray}
{\cal L}^{\hat a}_{g}=
\overline{\psi_Q} \bar\sigma^\mu T^{\hat a} \psi_{U} A^{\hat a}_\mu+
\overline{\psi_Q} \bar\sigma^\mu T^{\hat a} \psi_{E} A^{\hat a}_\mu+
\overline{\psi_L} \bar\sigma^\mu T^{\hat a} \psi_{D} A^{\hat a}_\mu+
{\rm h.c.}~~
\label{lg2}
\end{eqnarray}
In the above equations, $T^a$ and $T^{\hat a}$ correspond to the
unbroken and broken SU(5) generators, respectively, and 
$\bar\sigma_\mu=(1,-\sigma^k)$ with $\sigma^k$ being the Pauli
matrices. The effective 4-dimensional Lagrangian is given by:
\begin{eqnarray}
{\cal L}^{(4)}_g=\int dy \left[\delta(y)+\delta(-y+\pi R)\right] {\cal
L}_{g}(y)~.
\label{lg3}
\end{eqnarray}
The Yukawa interaction on the branes is given by 
\begin{eqnarray}
W^{(4)}=\int dy \left[\delta(y)+\delta(-y+\pi R)\right] W(y),~~{\rm
where}
\label{w1}
\end{eqnarray}
\begin{eqnarray}
W(y)=y_u~ 10~ 10~ H_5 + y_d~ 10~ {\bar 5}~ H_{\bar 5} + y_R~ 10~ {\bar
5}~ {\bar 5}.
\label{w2}
\end{eqnarray}
The $10~ 10~ H_5$ term contains the up quark mass term $Q U H_u^D$,
$10~ {\bar 5}~ H_{\bar 5}$ contains the down quark and electron mass
terms $Q D H_d^D$ and $L E H_d^D$ respectively, and $10~ {\bar 5}~
{\bar 5}$ can be decomposed as $Q L D + L L E + U D D$. Now comes the
task of assigning the matter $P'$ parities. The following three
observations have been made in \cite{altfer}:
\begin{enumerate}
\item
$P'(Q, L, U, D, E) = (+, +, +, +, +)$: After $y$-integration in
Eq.~(\ref{lg3}), the contribution from Eq.~(\ref{lg1}) survives and
that from Eq.~(\ref{lg2}) vanishes. So the SU(5) gauge interaction is
broken to the SM gauge interaction. But the simultaneous presence of
the lepton and baryon number violating operators inside $10 ~\bar{5}
~\bar{5}$ mediates proton decay at a very rapid rate.

\item $P'(Q, L, U, D, E) = \pm(+, +, -, -, -), ~{\rm or}~ \pm(+, -,
-, +, -)$: Full SU(5) symmetry is preserved on the branes. But, in
the first case, the Yukawa interactions are $P'$ odd and vanish after
the $y$-integration, while in the second case, up-type quark masses
cannot be generated (since $10~ 10~ H_5$ is $P'$ odd) leading to a
unrealistic picture. 

\item $P'(Q, L, U, D, E) = (+, -, +, +, -)$: SU(5) invariance is lost.
The interactions only respect SU(3) $\times$ SU(2) $\times$ U(1).
But (a) the necessary terms for yielding matter masses, namely, $Q U
H_u^D + Q D H_d^D + L E H_d^D$ can be generated, (b) the first term in
Eq.~(\ref{lg2}), an essential ingredient for vector boson mediated
proton decay, is prohibited, (c) coloured Higgsino mediated proton
decay operators, contained in Eq.~(\ref{w2}) but not explicitly shown
here (see \cite{altfer}), is again forbidden, and, finally, (d) the
lepton number violating $L L E$ and $L Q D$ terms, inside $ 10~
\bar{5} ~\bar{5}$, are also disallowed. Note, even though the baryon
number violating $U D D$ operator is allowed, this alone, without the
assistance of any lepton number violating operator, cannot drive
proton decay. We also observe that with this choice of $P'$ parities,
the SU(5) invariance at the brane is abandoned, and the symmetry
corresponds to that of the residual SU(3) $\times$ SU(2) $\times$
U(1). Still, some specific SU(5) properties, like $m_b = m_\tau$, can
be realised.

\end{enumerate}

So the lesson is that the most convenient $(P,P')$ assignments of
matter fields are: $(Q, U, D) = (+, +)$, and $(L, E) = (+, -)$. The
4-dimensional superpotential that follows this assignment is
\begin{eqnarray}
W^{(4)} = 2 \int dy \delta(y) \left[y_u Q U H_u^D + y_d (Q D H_d^D +
L E H_d^D) + y_R U D D \right].
\label{wfinal}
\end{eqnarray}
The first three terms in Eq.~(\ref{wfinal}) are responsible for the
mass generation of up-quarks, down-quarks and charged leptons
respectively. The last term is baryon number violating, and we will
see how it provides a handle to test this model.  Even though terms
involving $H_u^T$ and $H_d^T$ arise separately (not explicitly shown
in Eq.~(\ref{wfinal})), there is no bilinear Higgs triplet mixing
term. In fact, with the parity assignments above, proton decay is
forbidden at all orders.

We find it appropriate to mention here that the assignments of $P'$
parities made in the KAF model are by no means the only possiblity.
In the scenario presented by Hall and Nomura \cite{halnom}, the $P'$
assignments correspond to the case (2) listed above. The reason that
in \cite{halnom} this choice works is that Hall and Nomura suggest
that in some cases (e.g. $y_u$) the same Yukawa coupling at $y = 0$
and $\pi R$ may have a relative sign. This may, indeed, be true if the
Yukawa couplings were functions of the 5th co-ordinate $y$ and this
latter circumstance can occur quite naturally in string theories where
the couplings arise as vacuum expectation values of moduli.  So if one
were to start from string-inspired considerations, it may seem that
the choice made by Hall and Nomura is justified.  In the KAF model,
the approach is to treat the couplings as independent of $y$, i.e. to
have no relative sign between the same Yukawa coupling at $y = 0$ and
$\pi R$. On the other hand, the Hall-Nomura approach would allow a
relative sign between the same Yukawa coupling at $y = 0$ and $\pi R$,
thus permitting the $P'$ odd brane operators survive after the $y$
integration.  This is precisely the point where Hall and Nomura
\cite{halnom} differ with Alterelli and Feruglio \cite{altfer}. While
in \cite{halnom} a SU(5) symmetric brane interaction at $y = 0$ could
be written down, in \cite{altfer} the symmetry on the $y = 0$ brane is
that of the residual SU(3) $\times$ SU(2) $\times$ U(1). Furthermore,
the choice of $P'$ parities in \cite{halnom} predicts $A_\mu^{\hat a}$
boson exchanged proton decay at a rate to be seen in the forthcoming
experiments, while the choice made in \cite{altfer} prohibits all
kinds of proton decay at all orders.  The other important difference
between \cite{altfer} and \cite{halnom} is that while the former
allows baryon number violating term in the 4-dimensional
superpotential, the latter contains $R$-parity as a discrete subgroup
thus justifying the absence of either the lepton number or the baryon
number violating interaction. 

Thus having noted down the principal differences between \cite{altfer}
and \cite{halnom}, we now go back to the formalism of \cite{altfer}
which we have called the KAF model. Even though it is a GUT model, the
conventional `smoking gun' signal of proton decay is absent at all
orders. Turning the argument around, if proton decay is discovered,
this model will be ruled out. But then how we can experimentally
verify this model? What are its observable consequences? We address
ourselves to precisely this question in the present work. We observe
that the baryon number violating $U D D$ term in Eq.~(\ref{wfinal})
provides the clue as to where to look for its traces.

As such it is not possible to make any concrete estimation of the size
of $y_R$ in the KAF model. The best we can do is to argue a purely
geometrical suppresion of the $y_u$ or $y_d$ term, that contains one
bulk field ($H_u^D$ or $H_d^D$) and two brane-localised matter fields,
as compared to the $y_R$ induced term which contains all three
brane-localised fields. The basic point is that while writing down the
4-dimensional effective Lagrangian involving a 5-dimensional bulk
field, the latter has to be Fourier decomposed into a tower of
Kaluza-Klein modes with a normalisation $1/\sqrt{l}$ due to
compactification, where $l=\pi R/2$. In fact each time a bulk field
appears in a 4-dimensional effective Lagrangian, it is accompanied by
a volume suppression factor $\epsilon \equiv 1/\sqrt{M_*l}$, where
$M_*$ is the UV cutoff scale for the effective theory which appears as
a result of the correct normalization to provide the zero modes of the
bulk field with appropriate 4-dimensional canonical dimensions.  For
the KAF model, this line of reasoning has the implication that the
usual Yukawa couplings of the brane-localised matter fields with the
bulk Higgs field are suppressed by ${\cal O}(\epsilon)$ at the GUT
scale as compared to the baryon number violating Yukawa couplings
which involve all three matter fields.

But in the absence of any flavour physics whatsoever on the brane in
the KAF model it is not possible to make any numerical estimate of the
size of this suppression. On the other hand, in order to admit any
flavour physics on the brane, different matter fields will have to be
allowed to propagate in different dimensions in the bulk. However, we
stated in the beginning that bulk matter does not respect SU(5) gauge
invariance in 5 dimensions. Still, as has been observed in
Refs.~\cite{halnom,hall2}, one can order another set of $5'$ and $10'$
and their mirror partners to achieve a consistent bulk matter
formalism. Although this amounts to a departure from the usual GUT
structure, one can still ensure gauge coupling unification. In any
case, this is the minimal alteration from the `traditional line'
needed to activate any flavour physics at all on the brane. Without
going into the details of \cite{hall2}, we simply take a very crude
but reasonable estimate the authors of \cite{hall2} have made, namely,
$\epsilon \sim 1/10$.  Even though in the KAF model there is no such
extra $5'$ and $10'$, the approximate size of the volume suppression
that we borrow from \cite{hall2} seems a reasonable one. This choice
allows us to make a general statement that in the KAF model a generic
baryon number violating coupling at $M_{\rm GUT}$ is about an order of
magnitude larger than a generic Yukawa coupling at the same scale. In
the following paragraph we try to make this statement more explicit.

Even though the KAF model does not discuss any flavour physics at all,
we observe a significant hierarchy among the experimental quark and
lepton masses at low energy anyway. To this end, we make the following
simplistic assumption that the baryon number violating couplings may
have a kind of generational hierarchy that is observed among the usual
matter-Higgs Yukawa couplings. We first write down the effective
4-dimensional baryon number violating superpotential with explicit
generation indices as $\lambda''_{ijk} U_i D_j D_k$ (with an
antisymmetry between $j$ and $k$) and the 4-dimensional quark-Higgs
superpotential as $h_d^{ij} Q_i H_d D_j + h_u^{ij} Q_i H_u U_j$, where
$H_u$ and $H_d$ correspond to the zero modes of $H_u^D$ and $H_d^D$
respectively. For simplicity, we assume only two types of $\lambda''$
couplings, those involving the maximum and minimum possible values of
$(ijk)$, with the following relation at the GUT scale:
\begin{eqnarray}
\lambda''_{323} \sim h_t/\epsilon, ~~
\lambda''_{112} \sim h_u/\epsilon.
\label{hierar}
\end{eqnarray}
As the experimental upper limits imply, the other $\lambda''_{ijk}$
couplings are presumably related to the light quark Yukawa couplings.
These assumptions are admittedly simplistic, but we will try to see
what these simple assumptions will lead us to. We make a remark here
that if the $\lambda''$ couplings sit on their experimental upper
limits, it turns out that $\lambda''_{323} \gg \lambda''_{112}$
\cite{rpreview}.

With the kind of hierarchy at GUT scale given by Eq.~(\ref{hierar}),
particularly with $\lambda''_{323}$ being an order of magnitude larger
than $h_t$ and more specifically both being non-perturbative at the
GUT scale, an interesting possibility arises when both are driven to
their fixed point values at the electroweak scale. This has been
considered in \cite{barger} (see also \cite{carwhi}), and the fixed
point solutions are obtained in terms of the gauge couplings at the
scale $M_Z$ as ($Y_t = h_t^2/4\pi$, $Y_3 = \lambda^{\prime\prime
2}_{323}/4\pi$):
\begin{eqnarray}
Y_t \simeq {1 \over 16}\biggl \lbrack 8 \alpha_3+9\alpha_2+{9 \over 5}
\alpha_1 \biggr\rbrack ,~~ Y_3 \simeq {1 \over 16}\biggl \lbrack {56
\over 3} \alpha_3-3\alpha_2+{23 \over 15} \alpha_1 \biggr\rbrack .
\end{eqnarray}
Using the values $\alpha_3 (M_Z) = 0.118$, $\alpha_2 (M_Z) = 0.0336$
and $\alpha_1 (M_Z) = 0.0167$, we obtain $m_t =$ 174 GeV (for $\tan
\beta > 5$) and $\lambda''_{323} = 1.3$ at the weak scale.

What is the impact of such a large $\lambda''_{323}$ on electroweak
observables? Certainly, this coupling will contribute to $Z \to
b\bar{b}$ and $s\bar{s}$ via quark-squark triangle graphs
\cite{bcs}. The $Z$ coupling to $q_R$ ($q = s,b$) will be modified by
an amount $\lambda^{\prime\prime 2}_{323} N_c
f(m_t^2/\tilde{m}^2)/16\pi^2$, where $N_c = 3$ and $\tilde{m}$ is an
average squark mass. The explicit expression of the function $f$ in
terms of the Passarino-Veltman two- ($B$) and three- ($C$) point
functions are given in \cite{bcs}. The best constraint on
$\lambda''_{323}$ comes from $R_l = \Gamma_{\rm had}/\Gamma_l$ (where
$R_l^{\rm SM} = 20.740$, $R_l^{\rm exp} = 20.767 \pm 0.025$), given by
$\lambda''_{323} \leq 0.83$ (at 2$\sigma$) for a common squark mass of
100 GeV. We have used the latest LEP Electroweak Working Group report
\cite{lepewwg} to obtain the above limit as an update of our previous
work \cite{bcs}. This implies that the fixed point value
$\lambda''_{323} \simeq 1.3$ can be accommodated by taking the squark
mass near 150 GeV or so. We make a remark in passing that with this
interaction it is difficult to sizably alter the forward-backward
asymmetry in the $b$ quark channel keeping consistency with the
hadronic width measurement at the same time.

Another probe of $\lambda''_{323}$ is the $t \bar t$ production at the
Tevatron \cite{grs}. In the presence of the baryon number violating
couplings, in addition to the SM process, $q \bar q \rightarrow t \bar
t$ mediated by $s$-channel gluon exchange, there are new $t$-channel
processes involving the exchange of a squark. This new process can
lead to a significant enhancement of the $t \bar t$
cross-section. This enhancement can be effectively probed in Run II of
the Tevatron because of the high-statistics that will be achieved in
Run II and also because $t \bar t$ production at the Tevatron energy
is dominated by $q \bar q$-initiated channel which will be affected by
the baryon number violating operators. With a value of
$\lambda''_{323} \sim 1.3$ as suggested by the above arguments, we
compute the integrated $t \bar t$ production cross-section for
$\sqrt{s}=2$ TeV. The SM value for the $t \bar t$ cross-section at
this energy is about 9.4 pb and assuming an integrated luminosity of 2
fb$^{-1}$ we expect about 18,500 SM events. Assuming purely
statistical errors, we then estimate the sensitivity of this process
to probe the baryon number violating channel. We find that squark
masses to values somewhat larger than 300 GeV at the 95\% C.L. level
can be probed for $\lambda''_{323} \sim 1$.

What about the $\lambda''$ couplings which involve lower generation
indices? Intuitively, if our approximation in Eq.~(\ref{hierar}) is
sensible, then these couplings will be an order of magnitude larger
than the lower generation quark Yukawa couplings. In view of the
hierarchies among quark masses, it is rather impossible to ascertain
any value for those $\lambda''$ couplings, but roughly we can expect
them to be in the range $10^{-5}$--$10^{-3}$ at the GUT scale and more
or less in the same range at the weak scale. Some of these couplings
will induce large neutron-antineutron ($n$-$\bar{n}$) oscillation or
drive double nucleon decay into two kaons at a rapid rate. The
existing bounds from $n$-$\bar{n}$ oscillation are:
\begin{eqnarray}
\lambda''_{113} \leq 5 \times 10^{-3},~
\lambda''_{312} \leq 1.5 \times 10^{-2},~
\lambda''_{313} \leq 2.0 \times 10^{-2},
\end{eqnarray}
for an average squark mass of 200 GeV. These bounds have been obtained
\cite{goitysher,changkeung} using the $n$-$\bar{n}$ oscillation time
$\tau > 1.2 \times 10^{8} s$: there are some uncertainties though
coming from the nuclear matrix element calculations. The bound that
follows from the consideration of $NN \to KK$ is
\cite{goitysher,changkeung}:
\begin{eqnarray}
\lambda''_{112} \leq 1.0 \times 10^{-6},
\end{eqnarray}
again for a squark mass around 200 GeV. This bound also depends on a
suitable choice of the ratio between the hadronic scale and the SUSY
breaking scale. The values of $\lambda''$ couplings close to their
experimental upper limits can be easily accommodated in the KAF
model. Turning the argument around, if we observe a significant
enhancement (compared to the SM) in those channels, this might be
interpreted in terms of the $U D D$ couplings. The other lesson we
learn is that if we attempt to construct a realistic flavour model,
apart from the $\lambda''_{323}$, all other $\lambda''$ couplings
would have to be small (see the bounds listed in \cite{rpreview}).

In conclusion, the simplest version of the supersymmetric SU(5) GUT
models with an extra dimension compactified on a $S^1/(Z_2 \times
Z_2')$ orbifold, as introduced by Kawamura and further developed by
Altarelli and Feruglio, solves the doublet-triplet splitting problem
and suppresses proton decay at all orders by a suitable implementation
of matter $Z_2$ and $Z_2'$ parities. Since proton decay constitutes
the `smoking gun' signal of the Grand Unified models, a question that
naturally arises is how we experimentally verify this particular
incarnation of Grand Unification. Interestingly, this model admits
baryon number violating couplings, and in this paper we seek to
exploit the effects of these couplings on neutron-antineutron
oscillations, double nucleon decay into two kaons, hadronic decay
widths of the $Z$ boson, and $t \bar t$ production cross-section in
Run II of the Tevatron. Although a significant deviation from the SM
expectations on these observables will strengthen the case for a
baryon number violating interaction, any future observation of proton
decay will suffice to rule out the KAF model.

{\bf Acknowledgements:} It is a pleasure to thank G. Altarelli,
F. Feruglio, J. March-Russell and Y. Nomura for useful discussions. GB
thanks the Theory Division, CERN, Geneva, and ICTP, Trieste, for
hospitality during the period in which most of the work has been done.



\begin{thebibliography}{99}

\bibitem{susy} S. Dimopoulos, H. Georgi, Nucl. Phys. B 193 (1981) 150;
N. Sakai, Z. Phys. C 11 (1981) 153; R.K. Kaul, Phys. Lett. 109 B
(1982) 19; R.K. Kaul, P. Majumdar, Nucl. Phys. B 199 (1982) 36.

\bibitem{susygut} J. Ellis, D.V. Nanopoulos, S. Rudaz, Nucl. Phys. B
202 (1982) 43; P. Nath, A.H. Chamseddine, R. Arnowitt, Phys. Rev. D 32
(1985) 2348.

\bibitem{unifi} J. Ellis, S. Kelley, D.V. Nanopoulos, Phys. Lett. B
260 (1991) 131; U. Amaldi, W. de Boer, H. Furstenau, Phys. Lett. B 260
(1991) 447.

\bibitem{murayama} H. Murayama, A. Pierce, Phys. Rev. D 65 (2002)
055009 [hep-ph/0108104];
G. Altarelli, F. Feruglio, I. Masina, JHEP 0011 (2000) 040
[hep-ph/0007254].

\bibitem{sk} Y. Hayato {\em et al.} [Super-Kamiokande Collaboration],
Phys. Rev. Lett. 83 (1999) 1529 [hep-ex/9904020].

\bibitem{kawamura} Y. Kawamura, Prog. Theor. Phys. 105 (2001) 999
[hep-ph/0012125].

\bibitem{altfer} G. Altarelli, F. Feruglio, Phys. Lett. B 511 (2001)
257 [hep-ph/0102301].

\bibitem{koba} A.B. Kobakhidze, Phys. Lett. B 514 (2001) 131
[hep-ph/0102323].    

\bibitem{halnom} L. Hall, Y. Nomura, Phys. Rev D 64 (2001) 055003
[hep-ph/0103125].

\bibitem{hall2} L. Hall, J. March-Russell, T. Okui, D. Smith,
hep-ph/0108161. See also, A. Hebecker, J. March-Russell, Nucl. Phys. B
613 (2001) 3 [hep-ph/0106166]; A. Hebecker, J. March-Russell,
hep-ph/0107039.

\bibitem{rpreview} G. Bhattacharyya, Nucl. Phys. Proc. Suppl. 52A
(1997) 83 [hep-ph/9608415]; hep-ph/9709395; H. Dreiner,
hep-ph/9707435; B.C. Allanach, A. Dedes, H. Dreiner,
Phys. Rev. D 60 (1999) 075014 [hep-ph/9906209].   

\bibitem{barger} V. Barger, M.S. Berger, R.J.N. Phillips, T. Wohrmann,
Phys. Rev. D 53 (1996) 6407 [hep-ph/9511473].

\bibitem{carwhi} B. de Carlos, P.L. White, Phys. Rev. D 54 (1996) 3427
[hep-ph/9602381]; B. Brahmachari, P. Roy, Phys. Rev. D 50 (1994) 39,
Erratum-ibid. D 51 (1995) 3974.

\bibitem{bcs} G. Bhattacharyya, D. Choudhury, K. Sridhar,
Phys. Lett. B 355 (1995) 193 [hep-ph/9504314]. 

\bibitem{lepewwg} LEP Electroweak Working Group Report,  LEPEWWG Note
2001-01: Status of winter 2001. 

\bibitem{grs} D. Ghosh, S. Raychaudhuri, K. Sridhar, Phys. Lett. B 
396 (1997) 177 [hep-ph/9608352]. 

\bibitem{goitysher} J.L. Goity, M. Sher, Phys. Lett. B 346 (1995) 69,
Erratum-ibid. B 385 (1996) 500 [hep-ph/9412208]. For a more recent
analysis, see  K.S. Babu, R.N. Mohapatra, Phys. Lett. B 518 (2001) 269
[hep-ph/0108089]. 

\bibitem{changkeung} D. Chang. W.-Y. Keung, Phys. Lett. B 389 (1996)
294 [hep-ph/9608313]. 



\end{thebibliography}
\end{document}